\input epsf
\epsfverbosetrue
\documentstyle[twocolumn,aps,prb]{revtex}
\begin{document}
\draft
\title{Superconducting proximity effect in a mesoscopic ferromagnetic wire}
\author{M. Giroud$^1$, H.~Courtois$^1$, K. Hasselbach$^1$, D.~Mailly$^2$, and B.~Pannetier$^1$}
\address{$^1$ Centre de Recherches sur les Tr\`es Basses 
Temp\'eratures-C.N.R.S. associated to Universit\'e Joseph Fourier, 25 Ave. des Martyrs, 38042 Grenoble, France} 
\address{$^2$ Laboratoire de Microstructures et de Micro\'electronique-C.N.R.S., 196 Ave. H. Ravera, 92220 Bagneux, France}
\date{\today}
\maketitle
\begin{abstract}
We present an experimental study of the transport properties of a 
ferromagnetic metallic wire (Co) in metallic contact with a 
superconductor (Al). As the temperature is decreased below the Al 
superconducting transition, the Co resistance exhibits a significant 
dependence on both temperature and voltage. The differential 
resistance data show that the decay length for the proximity effect 
is much larger than we would simply expect from the exchange 
field of the ferromagnet.
\end{abstract}
\pacs{74.50.+r, 74.80.Fp, 85.30St}
\bigskip

Superconducting proximity effect consists in inducing 
superconductive properties in a  non-superconducting metal. 
Although this effect has been studied for a long time 
\cite{deGennes}, it has gained some renewed interest due to recent 
experiments performed on samples of mesoscopic size. In such 
samples, the electron phase-breaking length $L_\varphi$ is larger 
than the sample length $L$. One can thus probe experimentally the 
characteristic energy scale of the proximity effect $\epsilon_c = 
\hbar D / L^2$, which is the Thouless energy related to the sample 
length. This has led for instance to the observation of large 
magnetoresistance oscillations in normal metal (N) loops in contact 
with a superconducting (S) island 
\cite{Petrashov,Dimoulas,CourtoisPrl}. These oscillations provide a 
direct evidence for the long-range (up to $L_\varphi$) nature of the 
proximity effect. Another recent and striking result is the reentrant 
behaviour. The excess conductance induced by proximity effect is 
maximum at a temperature or a bias voltage equivalent to the 
sample Thouless energy \cite{CharlatPrl}, but the normal state 
conductance reappears at lower energy. 

Most experiments were performed in noble metals or 
semiconductor 2D electron gas, where electron interactions are 
negligible. In a free electron model, the zero-temperature, zero-bias 
resistance of a mesoscopic metallic wire is predicted to recover the 
normal state value \cite{Artemenko,StoofNazarov,Wilhelm}. In the presence of 
interactions, theoretical studies \cite{StoofNazarov,SpivakZhou} 
predict a severe modification of the transport properties. Attractive 
(respectively repulsive) electron-electron interactions are believed 
to result in a resistance lower (respectively higher) than the 
normal-state one \cite{StoofNazarov}. This could provide a probe 
for interactions in normal metals like Au, Ag, etc \cite{Mota}.

In this communication, we present an experimental study of the 
superconducting proximity effect in a ferromagnetic metal (F). 
Magnetic metals are in the strong interaction limit. Exchange 
interactions between electrons in a ferromagnet usually lead to 
efficient Cooper-pair breaking in F-S structures. However, it is 
worthwhile re-examining the actual proximity effect in a small 
ferromagnetic wire \cite{SpivakFerro}. Some experiments 
\cite{PetrashovFerro,Giordano} suggested long-range coherence 
effects, but without any clear conclusion.

\begin{figure}
\epsfxsize=6 cm 
\epsfbox{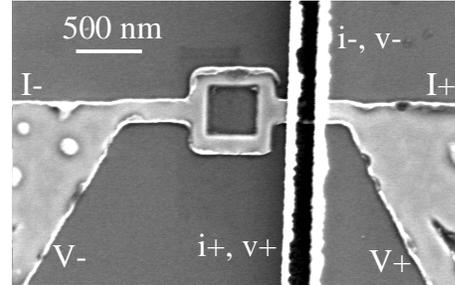}
\caption{Micrograph of Sample 1. The Co wire and loop is in 
metallic contact with one Al island. Thin residual Al strips appear 
on the sides of the Al wire. The side length of the Co loop is 500 nm. 
The four Co contact pads are labelled I+, I-, V+, V- . The Al island is 
patterned with four contacts i+, i-, v, +v-, as indicated. In sample 2 
(not illustrated), a second Al island is patterned on the left-hand 
side on the Co loop.}
\label{Photo}
\end{figure}

Samples (see Fig. 1) were fabricated using a two-step lift-off 
process. The 50 nm thick Co layer was e-beam evaporated on the 
patterned resist that was subsequently lifted off. The 100 nm thick 
Al islands were deposited after a soft in-situ ion-milling of the Co 
surface. The in-situ cleaning is a crucial step to achieve the desired high transparency of the Co-Al interface. In order to generate interferences, the Co conductor 
included a $0.5 \mu m$ square loop. The distance between the Co 
reservoirs is $2 \mu m$. Many samples were patterned on the 
same substrate, with zero, one or two Al islands. In the last two 
cases, one Al island was also linked to four contacts, in order to 
measure the Al wire and the Co/Al junction resistances. The width 
of the Co and Al wires were 100 and 140 nm respectively. Here we 
will focus on two typical samples labelled 1 and 2, with one and two 
Al islands respectively. The behaviour of each of these two samples is representative of the properties of four samples we measured.

Figure 2 shows the temperature dependence of the resistance of 
samples 1 and 2. The normal-state resistance of samples 1 and 2 is 
$96.09 \Omega$ and $98.35 \Omega$ respectively. With a Fermi 
velocity of $1.9 . 10^6 m/s$ in Co, we get an elastic mean free path 
$l_e$ of $1.1 nm$ and a diffusion coefficient $D = v_F l_e /3$ of $6.9 
cm^2/s$. As the temperature is decreased below the Al 
superconducting transition, the resistance of both samples 
decreases, reaches a minimum around 0.9 K and then increases. The temperature for the resistance minima is slightly different in the two samples, we do not have a simple explanation for that. The 
total amplitude of the variations is about $0.3 \%$ in sample 1 
and $0.8 \%$ in sample 2. In both cases, the low-temperature 
saturation value of the resistance is {\it larger} than the normal-
state resistance.

\begin{figure}
\epsfxsize=8.3 cm 
\epsfbox{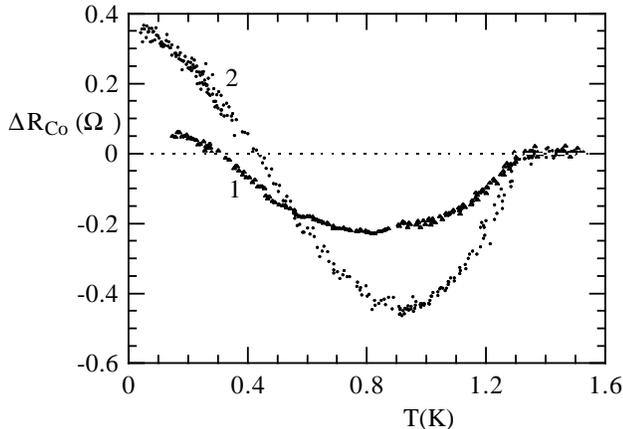}
\caption{Temperature dependence of the resistance of sample 1 
and 2. Sample 1 has one Al island in contact with the Co loop, 
sample 2 has two. The normal-state resistance, respectively $96.09 
\Omega$ and $98.35 \Omega$, has been subtracted. Bias current 
$0.1 \mu A$.}
\label{R(T)}
\end{figure}

\begin{figure}
\epsfxsize=8.3 cm
\epsfbox{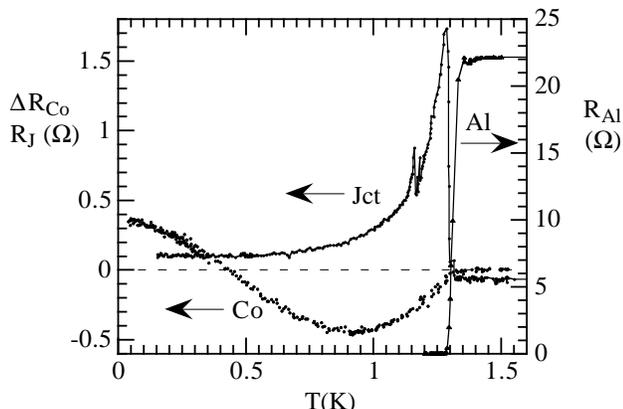}
\caption{Temperature dependence of the resistance of the Al wire 
(right hand scale), the Co wire and the Co/Al junction (left hand 
scale) of sample 2. Bias current $0.1 \mu A$.}
\label{R(T)CoAlJct}
\end{figure}

Figure 3 shows the same data for sample 2, together with the 
resistance of its Al wire and of the Co/Al junction. The Al wire 
becomes superconducting at 1.34 K, with a transition width of about 
10 mK. The superconducting properties of the Al wire are not 
strongly depressed by the proximity of the Co interface.

We measured the junction resistance by injecting current from one 
side (I+) of the Co wire to one side of the Al wire (i-) and measuring 
the voltage drop between the opposite sides of the Co and Al wires 
(V- and v+). The small negative offset above 1.34 K, when the Al wire is normal, stems
from a 3-dimensional spreading of the current lines in the metallic
electrodes of the junction. Such a sign reversal in a crossed-shaped
junction only occurs when the resistance of the junction is significantly smaller
than the electrodes resistances (here, $10 \Omega$ and $0.4 \Omega$ respectively for
Co and Al). This argument together with the measured resistance of $0.1 \Omega$
at the lowest temperature when the Al wire is superconducting confirms
that our junction is metallic. This order of magnitude is consistent with a
transparency $t$ of a few $\%$ after the relation $R_t^{-1} = 2 N(E_F) v_F S e^2 t$.
We believe that the junction resistance peak below 1.34 K is 
related to charge-imbalance effects in the Al island, when the gap is 
small compared to the injected quasiparticle energy. The Co 
resistance change is not much larger than the Al junction one, but 
we stress that the Co resistance varies significantly below 0.5 K, 
whereas the junction resistance does not vary anymore. This clearly 
shows that the variation of Co resistance is not due to a current 
redistribution effect induced by variations of the junction resistance 
\cite{2D}. 

\begin{figure}
\epsfxsize=8.3 cm \epsfbox{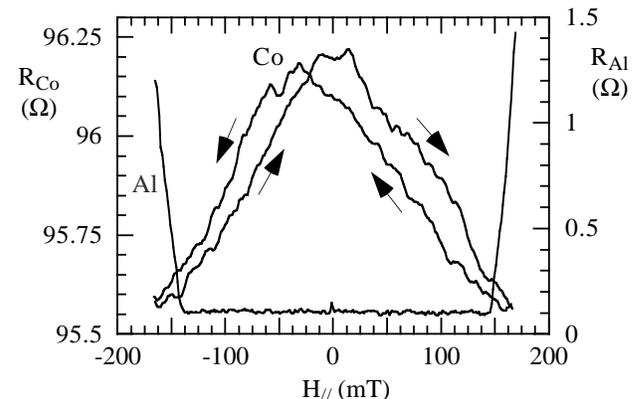}
\caption{Magnetoresistance of sample 1 Co wire and Al wire. The 
magnetic field is in the substrate plane, parallel to the Al wire. The 
Co resistance is hysteretic. The superconducting critical field of Al is 
close to 140 mT. Bias current $0.1 \mu A$.}
\label{MRPA}
\end{figure}

The magnetoresistance of sample 1 and 2 was studied in magnetic 
fields applied perpendicular or parallel to the substrate. Figure 4 
shows the magnetoresistance of sample 1 in parallel field. From the 
Al resistance, we measure a critical field of 140 mT for the Al wire. 
The Co wire has a small (less than $1 \%$ at 140 mT) and hysteretic 
magnetoresistance. No saturation is visible up to 170 mT. From this 
measurement, we can assert that our Co wire is ferromagnetic, but 
that all our experiments were performed in a regime where the Co 
magnetisation is not saturated. 

Figure 5 shows the magnetoresistance of the Co wire in 
perpendicular field, above the superconducting transition of Al at 
1.5 K, and well below at 0.29 K. There is no magnetoresistance at 
1.5 K, indicating that the Co magnetisation is in-plane. The 
perpendicular negative magnetoresistance only appears when Al 
becomes superconducting, and is about 8 times smaller than in 
parallel field. We searched for periodic oscillations of 
magnetoresistance as a function of perpendicular field. We achieved 
a resistance resolution better than $10^{-5}$, corresponding to 
$10^{-3} e^2/h$, by averaging over a large number of scans. The 
absence of Aharonov-Bohm oscillations at this level is a strong 
indication that the phase-breaking length in Co is smaller than $0.3 
\mu m$. This casts some doubts on the possible observation of 
weak-localisation--like effects in ferromagnetic films 
\cite{Giordano}.

\begin{figure}
\epsfxsize=8.3 cm \epsfbox{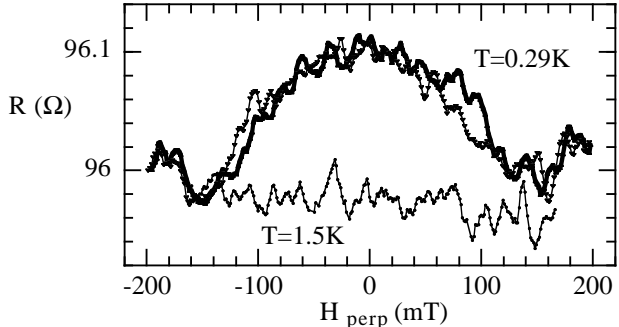}
\caption{Magnetoresistance of sample 1 at T = 0.29 K and T = 1.5 K 
in perpendicular field. Field scan from -200 to 200, then back to 
-200 mT. No periodic oscillations of the resistance were visible 
above the experimental noise. Bias current $0.1 \mu A$.}
\label{MRPE}
\end{figure}

We measured the differential resistance as a function of the bias 
current. Figure 6 shows three such curves recorded in different 
conditions. The upper curve was recorded at the lowest 
temperature 32 mK. The differential resistance exhibits a peak at 
zero bias, a minimum in the $1.7 \mu A$ range, and returns to the 
normal state value at high bias. This is strongly reminiscent of the 
re-entrance effect \cite{CharlatPrl}. We estimate that Joule heating 
did not exceed 0.3 K in the range below $5 \mu A$, and we checked 
that the junction resistance did not vary in this bias range. 

The lower two curves of Fig. 6 have been shifted for clarity. The 
middle curve was recorded at a temperature of 0.8 K. This curve 
looks very similar to the first one, with the remarkable exception 
that the zero-bias maximum is absent. In contrast, the high-bias 
features are unchanged. The lowest curve was recorded at very low 
temperature, in a parallel field of 130 mT, just below the Al critical 
field of 140 mT. This last curve shows a clear re-entrance peak of 
differential resistance below $1.0 \mu A$, but the high-bias peaks 
are no longer present. 

Let us compare these results with the ones previously reported for 
Cu \cite{CharlatPrl}. On increasing current, and thus voltage, we 
increase the energy of the electrons injected in the F-S mesoscopic 
sample. We can thus probe the energy dependence of the 
proximity-induced excess conductance. As in a "free electron" metal, 
we observe a resistance minimum both as a function of 
temperature and bias current. The differential resistance is 
maximum at zero temperature and voltage : this is the re-entrance 
effect for the metallic conductance of the normal metal.

\begin{figure}
\epsfxsize=8.3 cm \epsfbox{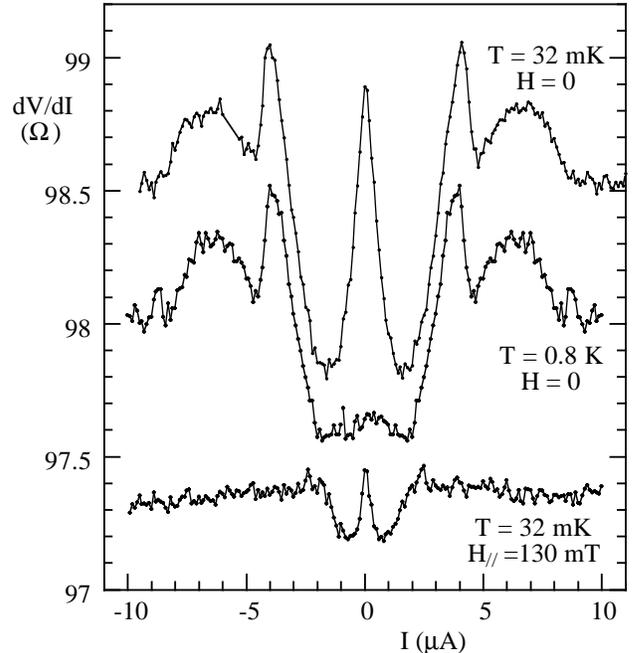}
\caption{Differential resistance of sample 2 measured in different 
conditions. Upper curve : T = 32 mK, H = 0. Middle curve : T = 0.8 K, 
H = 0. Lower curve : T = 32 mK, H = 130 mT (parallel to the 
substrate). The latter two curves have been offset by $0.5$ and $1 
\Omega$ respectively for clarity.}
\label{dV/dI(V)}
\end{figure}

The high-bias peaks are related to the Al superconducting gap 
and/or critical current. At T=32mK, the 130 mT magnetic field 
strongly depresses the Al gap, but does not affect the characteristic 
energies of the electrons injected in the F-S sample. Consequently, 
the re-entrance peak of the differential resistance at zero energy is 
still visible, but the high bias peaks disappear. On the other hand, at 
0.8 K and zero field, most of the electrons have an energy above the 
characteristic energy of the re-entrance effect, but the Al gap is not 
yet depressed, so that only the zero bias maximum disappears. This 
is consistent with our picture of the proximity effect.

In the quasiclassical theory, the temperature of the resistance 
minimum is $5 \epsilon_c / k_B$. The temperature dependence 
data gives us a Thouless energy of $14 \mu eV$ for the sample 2. With our 
estimate of the diffusion coefficient in Co, this energy would give a 
characteristic length of 180 nm, much shorter than the total sample 
length of $2 \mu m$. A simple interpretation of this result is that 
the Co electrons reflected on the Al island keep their phase 
coherence only on this shorter length scale. The effective 
mesoscopic sample length we are probing is only 180 nm. It is also 
the order of magnitude of magnetic domain sizes for Co samples 
deposited in similar conditions \cite{Wolfgang}. Indeed, it has been 
suggested that domain walls could contribute to decoherence of 
electrons \cite{Tatara}.

This decay length is much larger than the "exchange length" : 
$L_{exch} = \sqrt{ \hbar D / k_B T_{Curie}} = 2 nm$ in the dirty 
limit ($L_{exch}>l_{p}$), with $T_{Curie}=1388 K$ being the Curie 
temperature of Co. This length scale arises from the magnetic 
energy splitting between the incident electron and the Andreev
reflected hole in the exchange field of F \cite{Beenakker,Beasley}. 
In the Andreev reflection process, the electron spin is reversed. In 
consequence, the reflected hole has a different energy and 
momentum than the incident electron. This results in a finite decay 
length $L_{exch}$. 

If we take sample 2 normal state resistance ($98.35 \Omega$) to 
convert the current bias into a voltage corresponding to the 
minimum differential resistance, we get $170 \mu eV$. This is 
about 2.4 times larger than the voltage derived from the 
temperature dependence (Fig. 2), and even larger than the Al gap. 
This confirms that the coherence effects only occur on a length scale 
shorter than the total wire length. In comparison, 180 nm of our Co 
wire would have a resistance of about $8.9 \Omega$. At the current 
bias $1.7 \mu A$ of the minimum differential resistance, the 
voltage drop along this 180 nm coherence length is $15 \mu V$. 
This latter value is close to the Thouless energy derived from Fig. 2.

If we want to carry out a thorough quantitative analysis of our 
results, we encounter several difficulties : (i) A part of the 
resistance drop below the Al superconducting transition should 
originate from the local short-circuit by the Al island. A complete 
description would require taking into account the current 
redistribution in the Co thickness beneath the Al wire. (ii) The 
amplitude of the resistance drop is relatively small in comparison 
with the expected $15 \%$ variation for the resistance of the 
regions affected by the proximity effect. This could be related to the 
fact that we do not have good reservoirs injecting at a given energy 
from a well-defined distance, but diffusively distributed phase
breaking and inelastic events along the F wire.

Let us discuss the possible origin of the excess resistance, above the 
normal state residual resistance, at low temperature. This result is 
in agreement with the theoretical prediction of Stoof and Nazarov, 
that the zero-temperature and zero-voltage resistance of a N-S 
structure should exceed the normal-state resistance if repulsive e-e 
interactions are present. Following this viewpoint, our experiment 
would reflect a direct influence of the e-e interaction on a metallic 
resistance. From our data, we could extract a value for the
electron-electron interaction parameter. Another quite different explanation 
for the excess resistance at low temperature could be the screening 
of magnetic field by the superconductor in contact with the 
ferromagnet. The modification of the magnetic domain configuration 
near the F-S interface \cite{Buzdin} could enhance the resistivity in 
this region.

In conclusion, we have observed a proximity effect on the 
dissipative transport in a ferromagnetic metal in contact with a 
superconductor. Our results are in agreement with the early works 
of Petrashov \cite{PetrashovFerro} and Lawrence-Giordano 
\cite{Giordano}. From this work, we can assert than the behaviour 
described in Ref. \cite{Giordano} is due to a superconducting 
proximity effect in the ferromagnetic metal. The energy 
dependence of the effect has been probed through the temperature 
and voltage dependence of the resistance. The decay length for the 
coherence effect appears to be about 180 nm in the Co film. This 
value is of the order of the expected size of the magnetic domains in 
such films. The excess resistance, above the normal-state value, at 
zero voltage and temperature, could be explained by interaction 
effects.

We thank B. Spivak, P. Butaud and J. Caulet for stimulating discussions, D. 
Lafont, D. Mariolle (CEA-LETI) and Th. Fournier for the SEM 
micrograph of the samples, and J. Gilchrist for proof-reading the manuscript.

\end{document}